\documentclass[traditabstract]{aa}
\usepackage{graphicx}
\usepackage{natbib}
\usepackage{txfonts}
\usepackage{epsfig}
\usepackage{times}
\begin{document}

\title{The interaction-driven starburst contribution to the cosmic star formation rate density}

   \author{A. Lamastra, N. Menci, F. Fiore, P. Santini}
   \offprints{alessandra.lamastra@oa-roma.inaf.it}
   \institute{INAF - Osservatorio Astronomico di Roma, via di Frascati 33, 00040 Monte Porzio Catone, Italy.}
   \date{Received ; Accepted }
   \abstract{ An increasing amount of observational evidence supports the notion that there are two modes of star formation: a quiescent mode in disk-like galaxies,  and a starburst mode, which is generally interpreted as driven by merging. Using a semi-analytic model of galaxy formation, we derive the relative contribution to the cosmic star formation rate density of quiescently starforming and starburst galaxies,  predicted under the assumption that starburst events are triggered by galaxy encounters (merging and fly-by kind) during their merging histories.  
 We show that, within this framework, quiescently starforming galaxies dominate the cosmic star formation rate density at all redshifts. The contribution of the burst-dominated starforming galaxies  increases with redshift, rising from $\lesssim$ 5 \% at low redshift ($ z\lesssim $0.1) to $ \sim $ 20\% at $ z \geq $ 5.  We estimated that the fraction of the final ($ z $=0) galaxy stellar mass  which is formed through the burst component of star formation is  $ \sim $10\% for $10^{10} {\rm M_{\odot}} \leq$M$_*\leq10^{11.5} {\rm M_{\odot}}$ . Starburst galaxies, selected according to their distance from the galaxy main sequence, account for $\sim $10\% of the star formation rate density  in the redshift interval 1.5$<z<$2.5, i.e. at the cosmic peak of the  star formation activity.

   }
\titlerunning{The interaction-driven model for starbursts}
 \maketitle

\section{Introduction}
How  galaxies build up their stellar mass is a central question in galaxy formation. Many previous works  have demonstrated that a substantial fraction (30\%-50\%) of the stellar mass formed between $z\sim 3$ and $z\sim 1$ \citep{Dickinson03, Fontana03, Fontana04, Fontana06, Glazebrook04, Drory04, Rudnick06, Papovich06, Yan06, Pozzetti07}. However, it is still unclear what is the major channel for galaxy growth during this period.
Two star formation modes have been identified: the first is  a ``quiescent'' \footnote{In this paper, we use the term quiescent to refer to starforming galaxies that are not in interaction. We do not mean non-starforming systems, as the term is used in some literature.}
mode in disk-like galaxies where the star formation is extended over the whole galaxy disk and occurs  on time scales of  $\sim (1-2)$ Gyr ; while the second is a compact, starburst mode, where the star formation tends to be dominated by the nuclear regions and the gas depletion time scales are significantly lower ($\sim 10^7-10^8$yr) than those in quiescently starforming galaxies.  This accelerated mode of star formation is likely  triggered by major mergers, as suggested by observational evidence and theoretical arguments.  Observationally, major mergers are associated with enhancements in star formation in local Ultra Luminous Infrared Galaxies  \citep[ULIRGs,][]{Sanders96, Elbaz07},  and  some submillimeter galaxies \citep[SMGs,][]{Tacconi08, Daddi07, Daddi09, Engel10} . On the theoretical side, the merger-driven scenario for starburst galaxies is supported by the results of high-resolution numerical simulations, which showed the effectiveness of galaxy major mergers in causing the loss of angular momentum of the gas in the galaxy disk, with the consequent trigger of starburst events into the central regions of the galaxy  \citep{Hernquist89, Barnes91, Barnes96, Mihos94, Mihos96};  and by the success of semi-analytic models (SAMs) of galaxy formation, which include this additional channel of star formation, in reproducing the early formation of stars in massive galaxies  \citep[e.g.][]{Somerville01, Menci04, Menci05}.\\
An effective tools to understand the relative contribution of the different star formation modes is provided by
the scaling relations connecting the star formation rate (SFR) with global galactic quantities, like the gas mass (the Schmidt-Kennicutt relation)
or the total stellar mass (M$_*$). In fact, the normalization and/or the scaling of the SFR with such quantities differs for starbursts and quiescently starforming galaxies.
As for the former, the extension of the well-established local relation \citep{Kennicutt98} to higher redshifts has shown that quiescently starforming galaxies follow the local Schmidt-Kennicutt relation, while the SFRs of starburst galaxies are typically one order of magnitude greater than expected from their projected gas surface density \citep{Daddi10, Genzel10}. As for the correlation between the SFR and M$_*$, it has recently been shown that quiescently starforming galaxies lie along a ``main sequence'' characterized
by a typical, redshift-dependent value of the specific star formation rate (SSFR=SFR/M$_*$), while the less numerous starburst population is characterized by larger values of the SSFR. The main sequence is observed over a large range of redshift, from $z\sim 0$ to $z\sim 6$, and it is characterized by a small scatter ($\sim$ 0.3 dex), by a normalization which increases by a factor of $\sim$20 from $z \simeq 0$ to $z\simeq 2$, and by a slope which is sensitive to the technique used to select the sample of starforming galaxies and to the procedure for measuring M$_*$ and SFR, with values ranging from 0.6 to 1 \citep{Brinchmann04, Noeske07, Elbaz07, Daddi07, Daddi09, Santini09, Salim07, Stark09, Gonzalez11}. \\ 
The tecniques used to measure the SFRs and the stellar masses in galaxies are based on different observational tracers. A fruitful approach to estimate the stellar mass is the SED fitting.  This technique is based on the comparison between a grid of spectral templates  computed from standard spectral synthesis models and  the available  multiwavelength photometry  \citep{Papovich01, Shapley01, Shapley05, Fontana06, Santini09}.    
However, such a method does not provide reliable star formation histories at high redshifts, where the uncertainties become larger due to the SFR-age-metallicity degeneracies. An alternative commonly used estimator of the SFR is the UV rest-frame band, where young and massive stars emit most of their light.
Here the drawback is that dust absorbs, reprocesses, and re-radiates UV photons at near-to-far IR wavelengths, so that reliability of UV luminosity as a SFR tracer depends on large and uncertain corrections relying upon the dust properties.  Since most of the energy radiated by newly formed stars is reprocessed by dust, in order to accurately derive the SFR it is necessary to determine the dust bolometric output. In the past, the total IR luminosity (from 8 to 1000 $\mu$m) was calculated by extrapolating observation in the mid-IR.  With the launch of the \textit{Herschel} Space Observatory \citep{Pilbratt10} it has now become possible to measure the far-IR luminosity of distant galaxies at wavelengths where the dust emission is known to peak. \\
A recent study based on  $Herschel$ observations of starforming galaxies at 1.5$<z<$2.5 indicates  that only $ \sim $2\% of massive (M$_*>10^{10} {\rm M_{\odot}}$)  galaxies in this sample have starburst nature and account for only $ \sim $10\% of the star formation rate density  (SFRD) at $z\sim 2$ \citep{Rodighiero11}. This finding represent a test case for  the merger-driven scenario for starburst galaxies, since in this framework the occurrence of the starburst events is driven by the galaxy merger rates which  are expected to increase with redshift. 
In this paper  we  investigate this issue  in the framework of a theoretical model of galaxy formation that includes a physical description of starburst triggered  by  galaxy interactions during their merging histories.  The latter are described through Monte Carlo realizations, and are connected to gas processes and star formation using a semi-analytic model of galaxy formation in a cosmological framework \citep{Menci04, Menci05, Menci06, Menci08}.  In the model the starburst events are triggered not only by galaxy major mergers (i.e. the fusion of  galaxies with comparable mass, $m \approx m'$) but also by minor mergers ($m\ll m'$) and by closer galaxy interactions which do not lead to bound merging (fly-by events). Minor mergers and fly-by events induce starburst with a lower efficiency compared to major mergers,  but they are more probable, so they contribute appreciably to  the cosmic star formation history.\\
A description of the SAM  is given in Section \ref{model} ; in Section \ref{results} we derive the predicted  SFR-M$_*$ relation at  1.5$<z<$2.5 and the  contribution to the cosmic SFRD of starbursts and quiescently starforming galaxies; discussions and conclusions follow in Sections \ref{discussion} and \ref{conclusion}.

\section{The Model}\label{model}
We use the SAM described in details in \cite{Menci04, Menci05, Menci06, Menci08},  which connects, within a cosmological framework,
the baryonic processes (gas cooling, star formation, supernova feedback) to the merging histories of the dark matter (DM) haloes.  AGN activities, triggered by galaxy interactions in common DM haloes, and the related feedback processes are also included.
Here we recall the basic points.

\subsection{The dark matter merging trees}
Galaxy formation and evolution is driven by the collapse and growth of DM haloes, which originate from the gravitational instability of  overdense
regions in the primordial DM density field. This is taken to be a random,
Gaussian  density field with Cold Dark Matter (CDM) power spectrum within the
``concordance cosmology'' \citep{Spergel07} for which we adopt round
parameters  $\Omega_{\Lambda}=0.7$, $\Omega_{0}=0.3$, baryonic density
$\Omega_b=0.04$ and Hubble constant (in units of 100 km/s/Mpc) $h=0.7$. The
normalization of the spectrum is taken to be $\sigma_8=0.9$ in terms of the variance
of the field smoothed over regions of 8 $h^{-1}$ Mpc.

As  cosmic time increases, larger and larger regions of the density field
collapse, and  eventually lead to the formation of groups and clusters of
galaxies; previously formed, galactic size  condensations are enclosed. 
The merging rates of the DM haloes are provided by the Extended Press \& Schechter formalism \citep[see][]{Bond91, Lacey93}. 
 The clumps included into larger DM haloes
may survive as satellites, or merge to form larger galaxies due to binary
aggregations,  or coalesce into the central dominant galaxy due to dynamical
friction; these processes take place over time scales that grow longer over
cosmic time, so the number of satellite galaxies increases as the DM host haloes
grow from groups to clusters \citep[see][]{Menci05, Menci06}.

\subsection{The star formation law}
\begin{figure*}[t!]
\begin{center}
\includegraphics[width=6 cm]{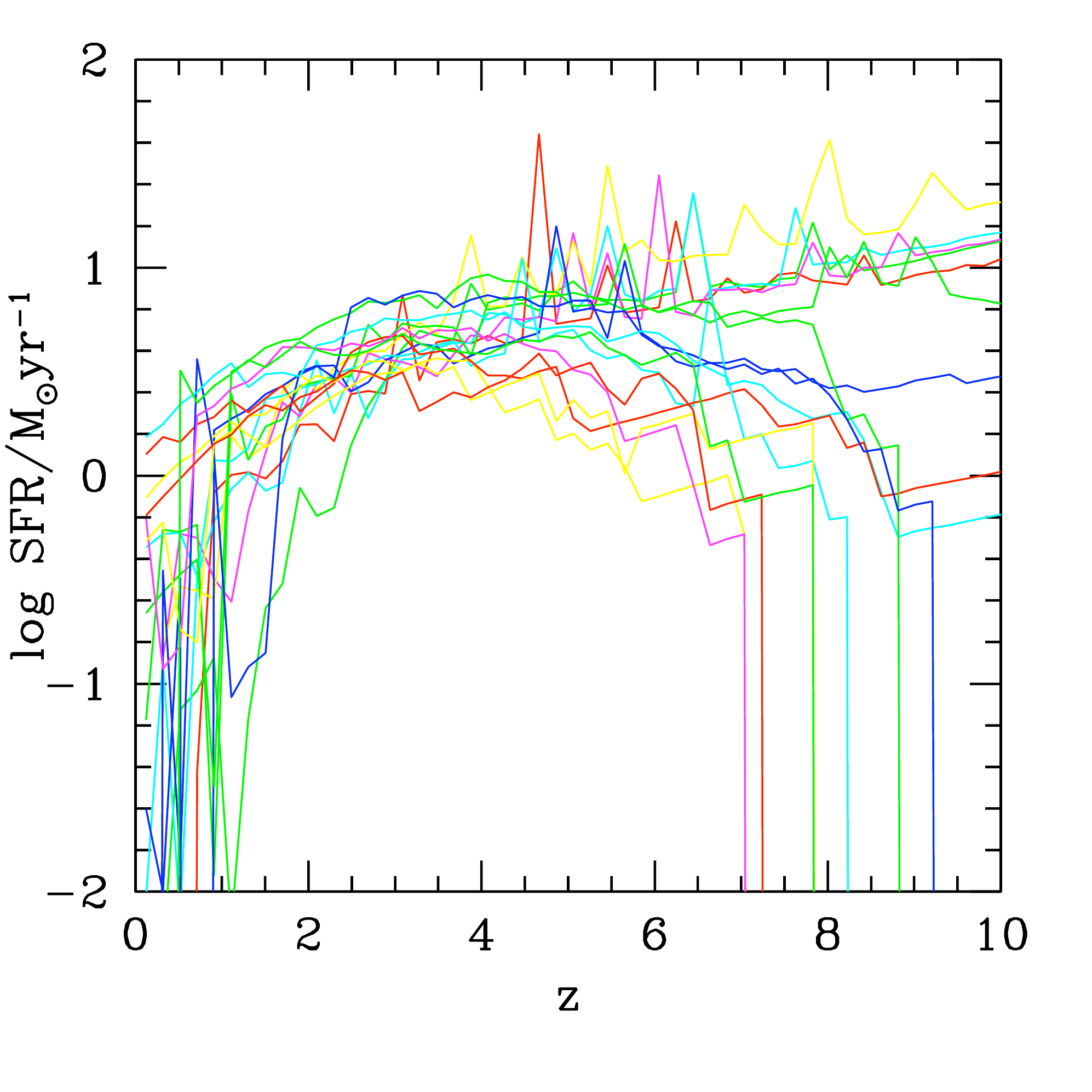}
\includegraphics[width=6 cm]{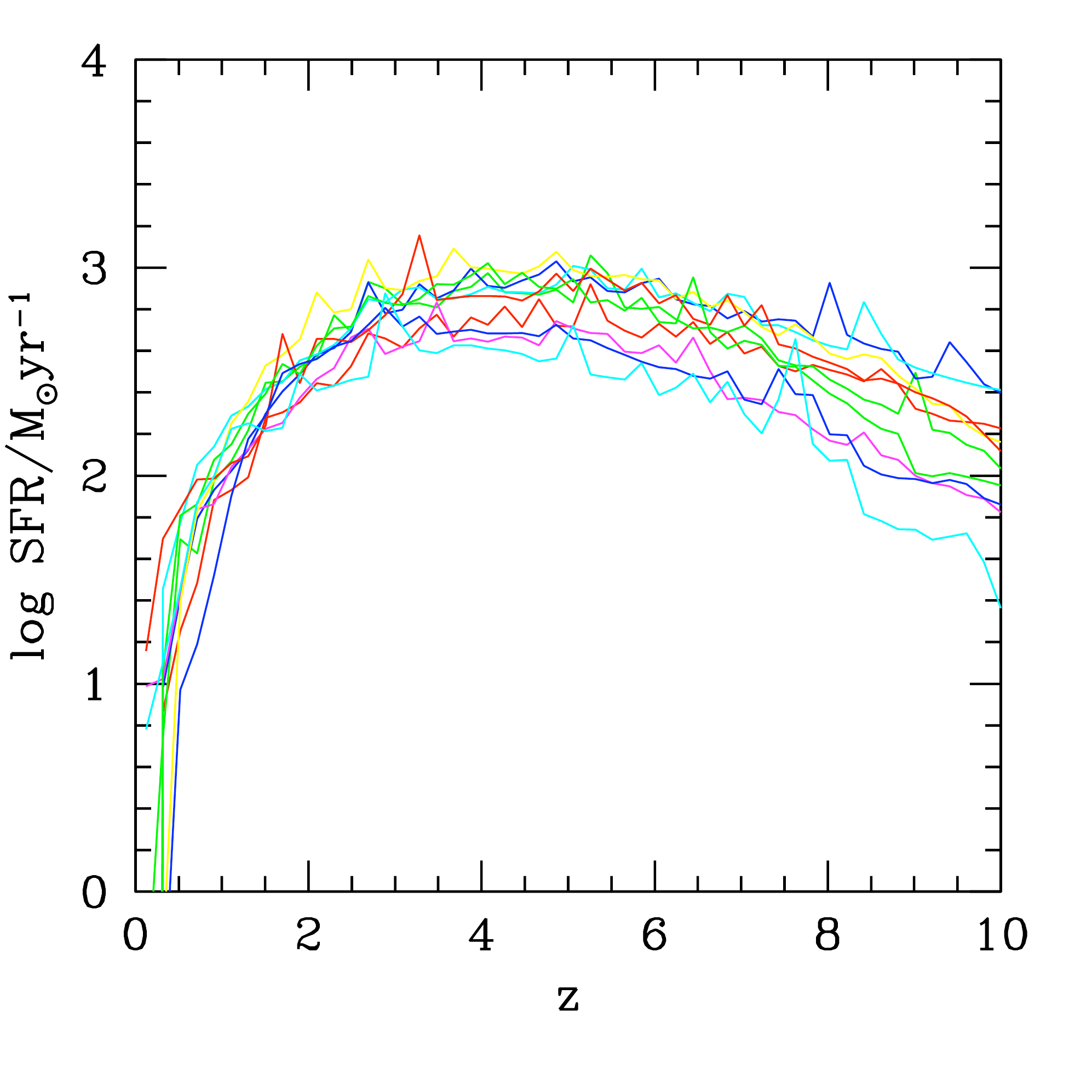}
\caption{SFHs drawn from the Monte Carlo realizations of galaxies that, at $ z=0 $ , have masses  of  M$_*\simeq10^{10} {\rm M_{\odot}}$ (left) and M$_*\simeq10^{12} {\rm M_{\odot}}$ (right). The SFHs are obtained by summing over all the progenitors that have merged to form  galaxies of the selected final mass. The quiescent star formation  is represented by the smooth component of the SFHs,  the burst component by the impulsive one. The duration of the SFR-excess phase due to the interactions is represented by the width of the peak of the SFHs.}
\label{SFH}
\end{center} 
\end{figure*}
The processes connecting the baryonic components to the growing DM 
haloes (with mass $m$) are computed as follows. Starting from an initial amount $m\Omega_b/\Omega$ of gas at virial temperature of the galactic haloes, we compute the mass $m_c$ of cold baryons within the cooling radius.
The cooled gas mass $m_c$ settles into a rotationally supported disc with radius $r_d$
(typically ranging from $1$ to $5 $ kpc), rotation velocity $v_d$
and dynamical time $t_d=r_d/v_d$, all computed after \cite{Mo98}. 
 Two channels of star formation 
may convert part of such  gas into stars: \newline
i) quiescent star formation, corresponding to the gradual conversion of the cold gas in the galaxy disk into
stars, for which we assume the canonical Schmidt-Kennicutt form:
\begin{equation}
 {\rm SFR_{q}} = m_c/\tau_q
 \label{sfr_q}
\end{equation}
where $\tau_q=qt_d$, and  $q$ is a model free parameter which is chosen to match the \cite{Kennicutt98}  relation; 
 
ii) burst-like star formation triggered by galaxy interactions, at a rate:
\begin{equation}
{\rm SFR_{b}} \propto f m_c/\tau_{b}
\label{sfr_b}
\end{equation}
here the time scale $\tau_{b}$  is assumed to be the crossing time  for the destabilized cold gas component ($t_d$), and $f$  is the fraction of cold gas destabilized by the encounters. A clear visualization of the two star formation modes implemented in the model is given in figure \ref{SFH}, where we show the star formation histories (SFHs) for two subsets of galaxies selected from the Monte Carlo simulations. 
The fraction $f$ of cold gas destabilized by the encounters has been worked out by  \cite{Cavaliere00}  in terms of the  variation $\Delta j$ of the specific angular momentum $j\approx
Gm/v_d$ of the gas to read  \citep{Menci04}:
\begin{equation}
f\approx \frac{1}{2}\,
\Big|{\Delta j\over j}\Big|=
\frac{1}{2}\Big\langle {m'\over m}\,{r_d\over b}\,{v_d\over V_{rel}}\Big\rangle\, 
\label{f}
\end{equation}
here $b$ is the impact parameter, evaluated as the greater of the radius $r_d$ and the average distance of the galaxies in the halo, $m'$ is the mass of the  partner galaxy in the
interaction, $V_{rel}$ is the relative velocity between galaxies, and the average runs over the probability of finding such a galaxy
in the same halo where the galaxy with mass $m$ is located.  The prefactor accounts for the probability 1/2 of inflow rather than outflow related to the sign of $\Delta j$.
We assume for the proportionality  constant in eq. (\ref{sfr_b}) the value of 3/4, 
while the remaining fraction of the inflow is assumed to feed the central black hole \citep[see][]{Sanders96}.   When applied to the black hole accretion and to the related AGN emission  the above model has proven to be very successful in reproducing the observed properties of the AGN population from $ z=6 $  to the present  \citep{Menci04, Menci08, Lamastra10}.   \\
 The probability for a given galaxy  to be in a burst phase is defined as the ratio $\tau_{b}/\tau_r$ of the duration of the burst to the average time interval between bursts.
The rate of such  encounters  $\tau_r^{-1}$ is 
\begin{equation}\label{int}
\tau_r^{-1}=n_T\,\Sigma (r_t,v,V_{rel})\,V_{rel}
\end{equation}
here $n_T=3 N_T/4\pi R^3$ is the number density of galaxies in the same halo,
$V_{rel}$ is their relative velocity, and $\Sigma \simeq \pi\langle r_t^2+r_t^{'2}\rangle$  is the cross section for
such encounters which is  given by  \cite{Saslaw85}  in terms of the tidal radius
$r_t$ associated to a galaxy with given circular velocity $v$ \citep[see][]{Menci04}. \\
At each time step, the mass $\Delta m_h$ returned from  the cold gas content of the disk to the  hot gas phase due to the energy released by SNae following star formation is estimated from canonical energy balance arguments as $\Delta m_h=E_{SN}\epsilon_0 \eta_0 \Delta m_* /v_c^2$, where $ \Delta m_*  $  is the stellar mass formed in the time step, $\eta_0$ is the number of SNe per unit solar mass (for a Salpeter initial mass function $\eta_0 = 6.5 \times 10^{-3} {\rm M_{\odot}}^{-1}$), $E_{SN}=10^{51} erg/s$ is the energy of ejecta of each SN, and $v_c$ is the circular velocity of the galactic halo; $\epsilon_0=0.01$ is a tunable efficiency for the  coupling  of the emitted energy with  the cold interstellar medium.  
At each merging event, the masses of the different baryonic phases ($\Delta m_c$, $\Delta m_*$ and $\Delta m_h$)  in each galaxy are refueled by those in the merging partners. Thus,  for each galaxy the star formation  is driven by the cooling rate of the hot gas and by the rate of refueling of the cold gas, which in turn is intimately related to the galaxy merging histories.   The model has been tested against several observed properties of the local galaxy population like the $B$ band luminosity function, the stellar mass function, the Tully-Fisher relation, the galaxy bimodal colour distribution,  the  
distribution of cold gas and disk sizes. At higher redshifts, the model has been tested against the evolution of the luminosity functions and the number counts in different bands \citep{Menci05, Menci06}.


\section{Results}\label{results}
\subsection{The SFR-M$_*$ relation}\label{sfr_m}
\begin{figure*}[t!]
\begin{center}
\includegraphics[width=8 cm]{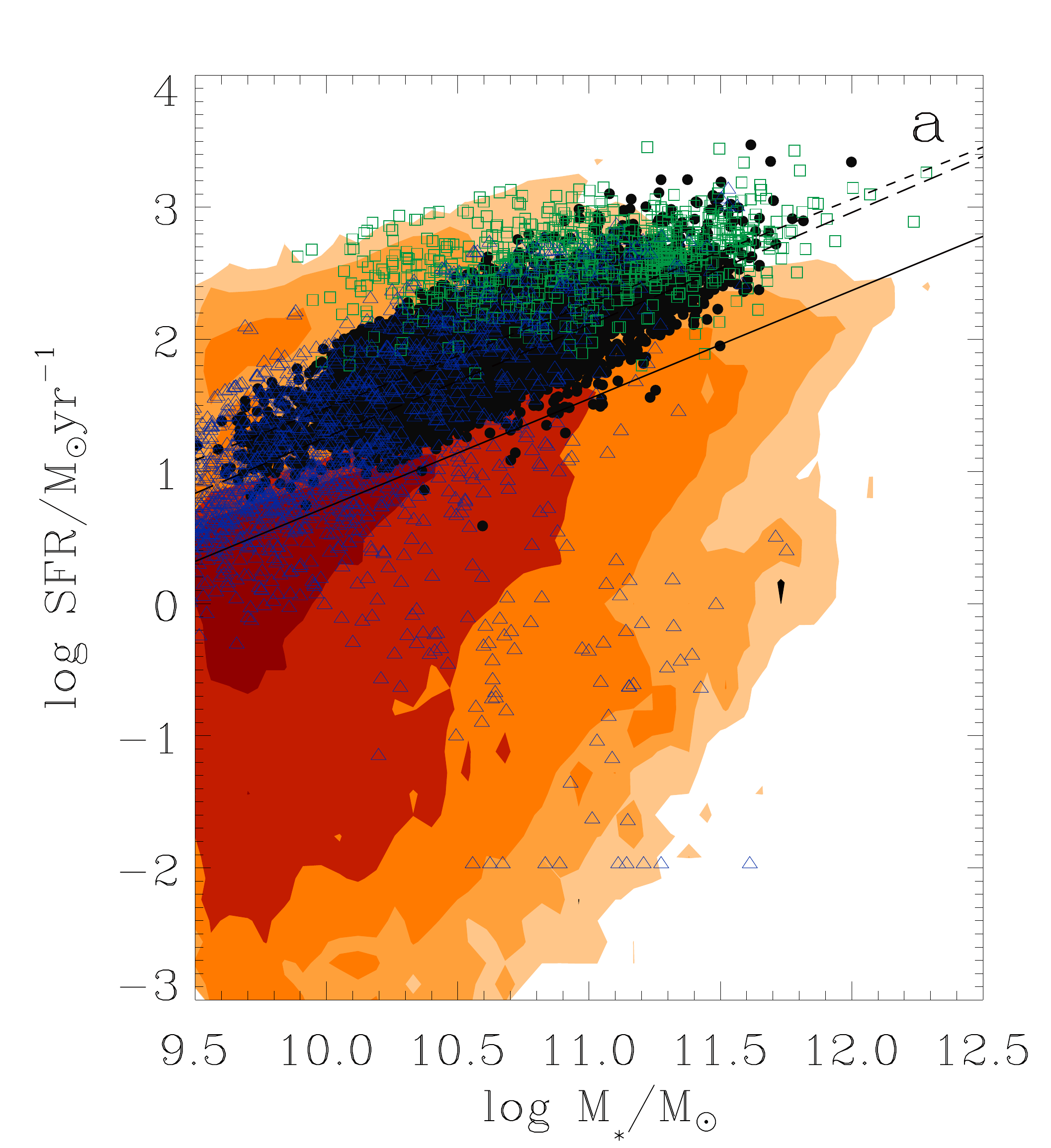}
\includegraphics[width=8 cm]{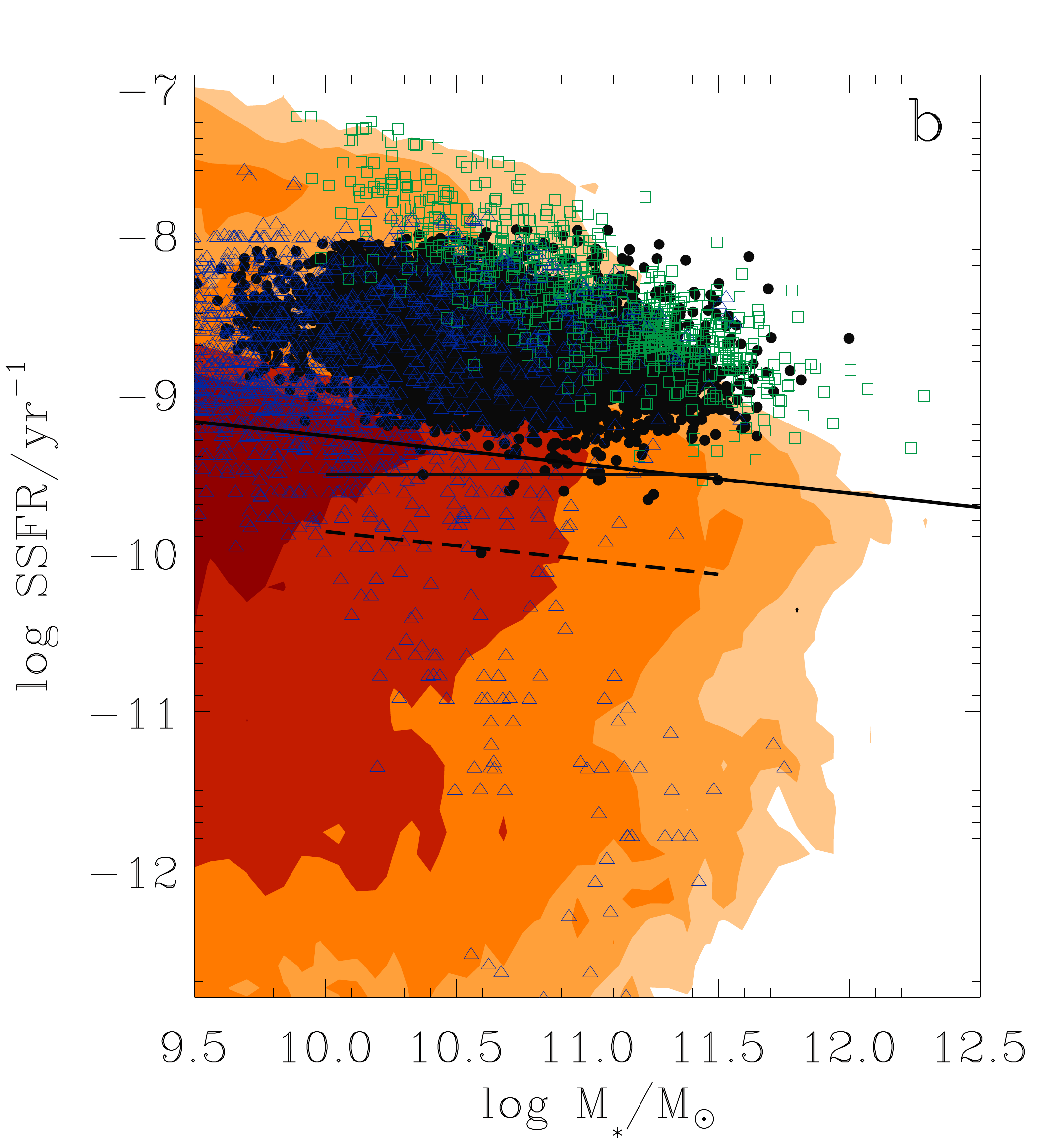}
\caption{Left: SFR-M$_*$  relation at $1.5<z<2.5$. The five filled contours correspond to equally spaced values of the density (per Mpc$^3$) of model galaxies in logarithmic scale: from 10$^{-6}$  for the lightest filled region to  10$^{-2}$ for the darkest.  Green squares: $Herscel$-PACS- selected  galaxies, black circles: BzK galaxies, from the \cite{Rodighiero11} sample; blue triangles: GOODS-MUSIC catalog from the \cite{Santini09} sample. The dashed and long-dashed lines  indicate the best-fit of the galaxy main sequence obtained by \cite{Rodighiero11}  and \cite{Santini09} respectively, while the solid line  shows the spline fit to the peaks of the SFR  distributions of model galaxies as a function of stellar mass.
Right: SSFR as a function of M$_*$. Colour code, symbols  and solid line as in the left panel. The dashed and thin solid lines indicate the ${\rm SSFR>SSFR_{MS}}$-2$\sigma$ and the SSFR$ > t_H(z)^{-1}$ limits 
of the galaxy samples used in the analysis of Section \ref{res2} respectively.}
\label{sfr_mst}
\end{center} 
\end{figure*}
\begin{figure*}[t!]
\begin{center}
\includegraphics[width=8  cm]{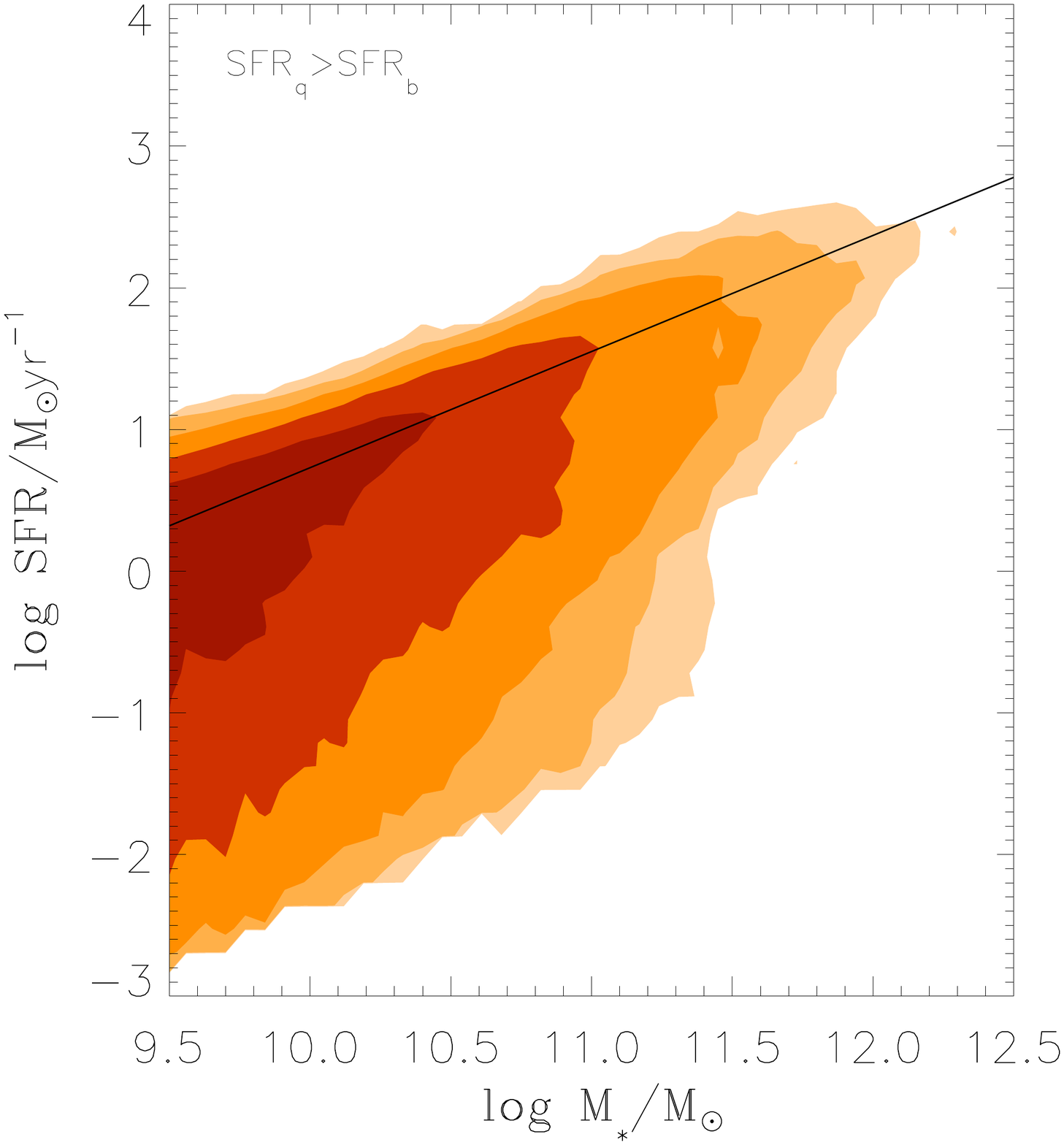}
\includegraphics[width=8  cm]{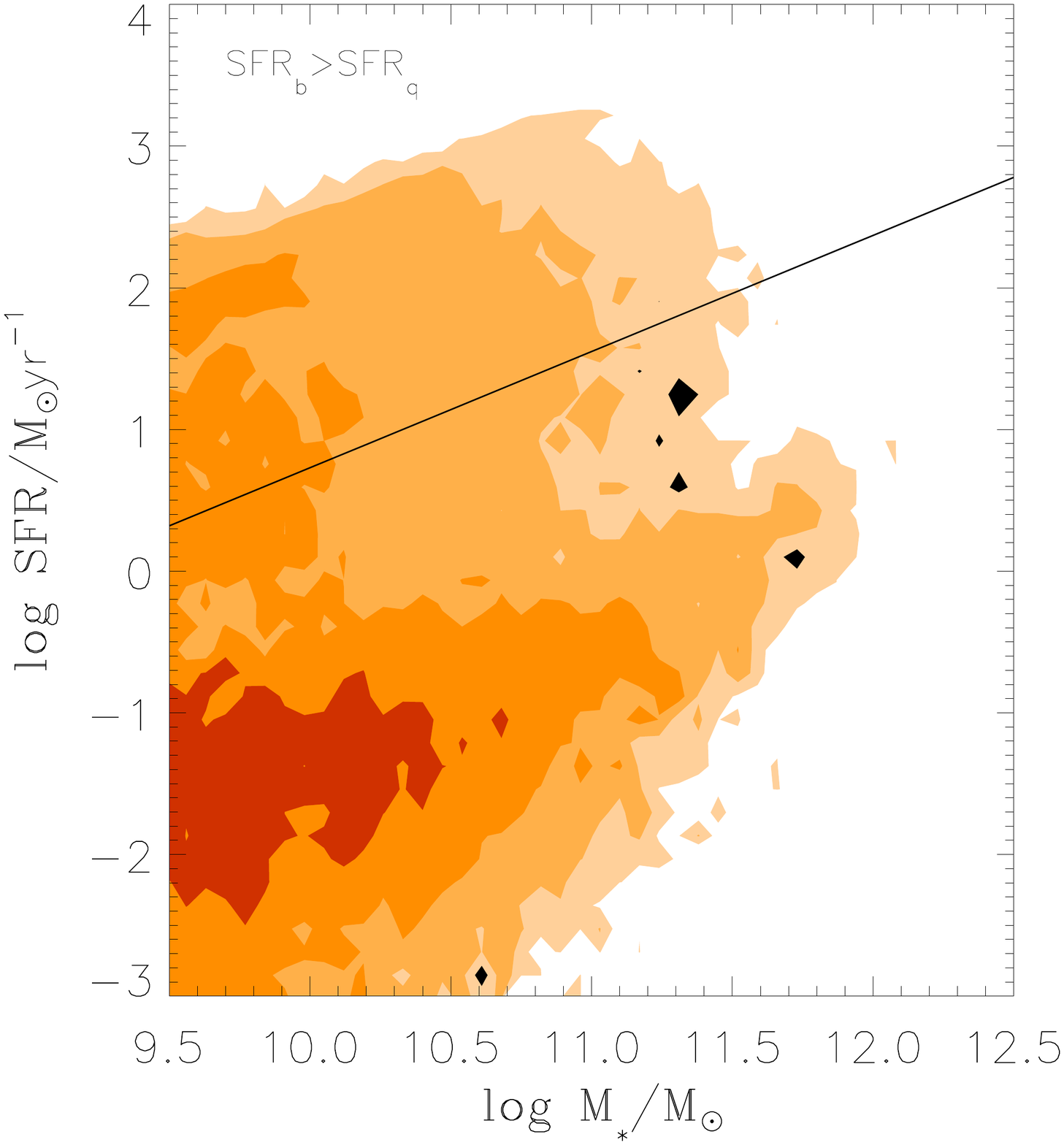}
\caption{The predicted SFR-M$_*$ relation at $1.5<z<2.5$ obtained selecting from the model galaxies dominated by the quiescent (left) and burst (right) mode of star formation. Contours and solid line as in figure \ref{sfr_mst}.}
\label{sfr_mst_SF}
\end{center} 
\end{figure*}

As a first step in the study of the  contribution of starbursts and quiescently starforming galaxies to the cosmic SFRD, we compare  in figure \ref{sfr_mst}  the predicted SFR-M$_*$  (left)  and  SSFR-M$_*$ (right) relations for model galaxies  with M$_* > 10^{9.5}{\rm M_{\odot}}$ at $1.5<z<2.5$ (coloured contours) with those obtained by \cite{Rodighiero11}  and by \cite{Santini09}. The former authors used a combination of  BzK- and $Herscel$-PACS- selected starforming galaxies  from the COSMOS and GOODS fields.  The BzK colour selection is designed to identify starforming galaxies in the redshift interval $1.5<z<2.5$ by using colours that samples key features in the spectral energy distributions of galaxies, mainly the rest-frame 4000 $ \AA $  break and the UV continuum slope  (BzK$ \equiv (z-K)_{AB}-(B-z)_{AB}\geq$-0.2, \citealt{Daddi04}). The SFRs of BzK galaxies are estimated from the UV-rest frame luminosity corrected for dust reddening and they yield  an almost linear correlation with the stellar mass (black circlces in fig.  \ref{sfr_mst}).
The  $Herschel$-PACS detection limit depends on  redshift;  at $ z\simeq 2$ it corresponds to $\sim$200 M$_{\odot}$/yr  and to $\sim$50 M$_{\odot}$/yr  for the COSMOS and GOODS fields respectively. The PACS-based SFRs run almost flat with the stellar mass and occupy the upper envelop of the distribution (green squares in fig.  \ref{sfr_mst}). This different behaviour could be ascribed to the different selection of the two samples, one being SFR limited, and one  being mass limited \citep{Rodighiero11}.
The \cite{Santini09}  sample is based on the  GOODS-MUSIC catalog; for dectected sources the SFRs are derived
from the 24$\mu$m emission, corrected to take into account the well-known overestimation of the SFR for bright
sources at $z \sim 2$ \citep[see][for further details]{Santini09}, while for undetected objects the SFRs are derived from  SED
fitting in the optical/near-IR range.  In contrast with the previous sample, this catalog includes also passive galaxies, and this explains the larger extention of the SFR-M$_*$ relation towards low star formation rates (blue triangles in fig.  \ref{sfr_mst}).  These data lie on the predicted confidence region represented by the contour plot, however the bottom-left part of the SFR-M$_*$  diagram remains unsampled due to observational incompleteness. This observational limitation does not  allow us to test one of the striking feature of the SFR-M$_*$ relation predicted by hierarchical models of galaxy formation, namely, the behaviour of the scatter. In fact,  its increase with decreasing stellar mass   
 is due to  the large variety of star formation  histories corresponding to the low-mass population, while the similarity of the star formation histories of high-mass galaxies results in smaller scatter in the SFR-M$_*$ plane.  This is especially true at high redshifts $z>6$; at such epoch, the small number of progenitors of present-day low-mass galaxies
(and their different collapse time) yield a large variance in the star formation histories, while the larger number of progenitors of high-mass galaxies
(already collapsed since they formed in biased regions of the density field) result in a smaller variance among the different star formation histories (see fig. \ref{SFH}). \\
The dashed and long-dashed lines  in figure \ref{sfr_mst} indicate the best-fit of the galaxy main sequence obtained by \cite{Rodighiero11}  and \cite{Santini09} respectively, while the solid line shows the spline fit to the peaks of the SFR (SSFR) distributions of model galaxies as a function of stellar mass (log SFR=0.8*logM$_*$-7.5).
The slopes of the observed and predicted SFR-M$_*$ relations are similar (0.8 $\leq\alpha \leq$0.9), while the normalization of the predicted relation is a factor of 4-5 lower than the observed relations.
 The offset between the predicted and observed SFR-M$_*$ relations in the local universe  \citep[$ z\simeq 0.1$][]{Brinchmann04}   reduces by a factor of about two, implying that the evolution of the SSFR with redshift predicted by the model is slower than that observed  at $ z\lesssim 2$. The local distribution of model galaxies in the  SFR-M$_*$   plane shows a bimodal behaviour,  with a  clear separation between actively star forming and passively evolving galaxies, as seen in observations \citep{Brinchmann04}.\\
 Similar offsets between the observed SFR-M$_*$  relation and those expected by various kind of SAMs and hydrodinamical simulations  has been previously reported in the literature \citep[see e.g.  ][]{Daddi07, Dave08, Fontanot09, Damen09, Santini09,Lin12}. 
Understanding whether this mismatch originates from  systematic effects in the stellar mass and/or SFR determinations, or from incompleteness in our basic picture of galaxy assembly is a very difficult task. The amplitude of this offset depends on the the technique  used to estimate the SFR and stellar mass and on the sample selection (see the offset between the dashed and long-dashed lines). 
In our analysis of the role of starbursts relative to quiescently starforming galaxies (section \ref{res2})  we shall normalize both the model and observed SFRs to their main sequence values, and we will address this mismatch in more detail in section \ref{discussion}.\\
To understand the role of starbursts and quiescently starforming galaxies in deriving the SFR-M$_*$ relation we separately show in figure \ref{sfr_mst_SF} the relations obtained selecting from the Monte Carlo simulations  galaxies dominated by the quiescent mode of star formation (${\rm SFR_{q}}>{\rm SFR_{b}}$, left panel), and galaxies dominated by the burst mode of star formation (${\rm SFR_{b}}>{\rm SFR_{q}}$, right panel).
The former show a tight correlation between the SFR and the stellar mass. This is a natural outcome, being this component a fairly steady function of time (see the smooth component of the SFHs in fig. \ref{SFH}).  The scatter of this relation originates from the distribution of cold gas at a given stellar mass, and ultimately stems from the stochastic nature of the merging histories of galaxies.  
 When we select galaxies where  ${\rm SFR_{b}}>{\rm SFR_{q}}$ we obtain a more scattered distribution  in the SFR-M$_*$ plane.  In these galaxies the star formation rate is not only determined by the amount of cold gas available in the galaxy disk, but it is also regulated by the amount of such gas which is destabilized during galaxy interactions (see eq. \ref{sfr_b}). Minor mergers and fly-by events, which dominate the statistics of encounters in the hierarchical clustering scenario,  induce a lower fraction of  destabilized gas compared to major mergers (note the dependence of $ f $ on  the mass ratio $ m'/m $ in eq. \ref{f}).  This analysis show that the main sequence is mainly populated by quiescently starforming galaxies, while the loci well above and below it by galaxies experiencing major mergers and minor mergers/fly-by events respectively.
 

\subsection{The contribution of starbursts and quiescently starforming  galaxies to the cosmic SFRD}\label{res2}
\begin{figure}[h!]
\begin{center}
\includegraphics[width=9cm]{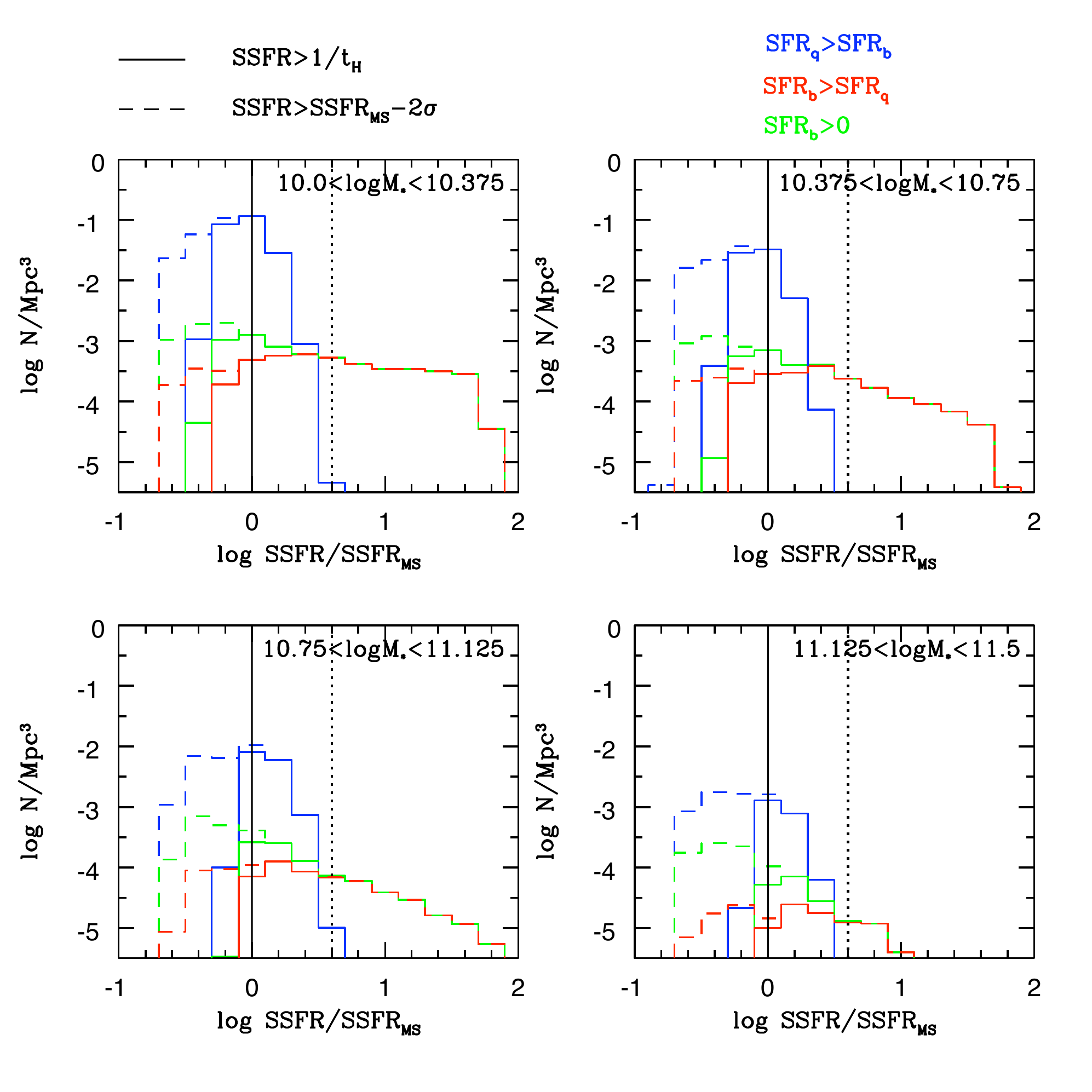}
\caption{SSFR distributions of model galaxies, at  $1.5<z<2.5$, in four mass bins. The blue histograms refer to galaxies dominated by the quiescent mode of star formation, the  red histograms to  galaxies dominated by the burst mode of star formation, and the green histograms to galaxies with an interaction-driven star formation component. The solid vertical lines show the position of the main sequence, while the  vertical dotted lines show the SSFR threshold used to identify starburst galaxies: ${\rm SSFR_{SB}}$ $> 4 \times$ ${\rm SSFR_{MS}}$. The solid  distributions are obtained selecting galaxies with  SSFR$>t_H(z)^{-1}$, while dashed distributions are obtained selecting galaxies with SSFR$>$ ${\rm SSFR_{MS}}$-0.6 dex.}\label{ssfr_dist}
\end{center} 
\end{figure}

\begin{figure*}[t!]
\begin{center}
\includegraphics[width=8cm]{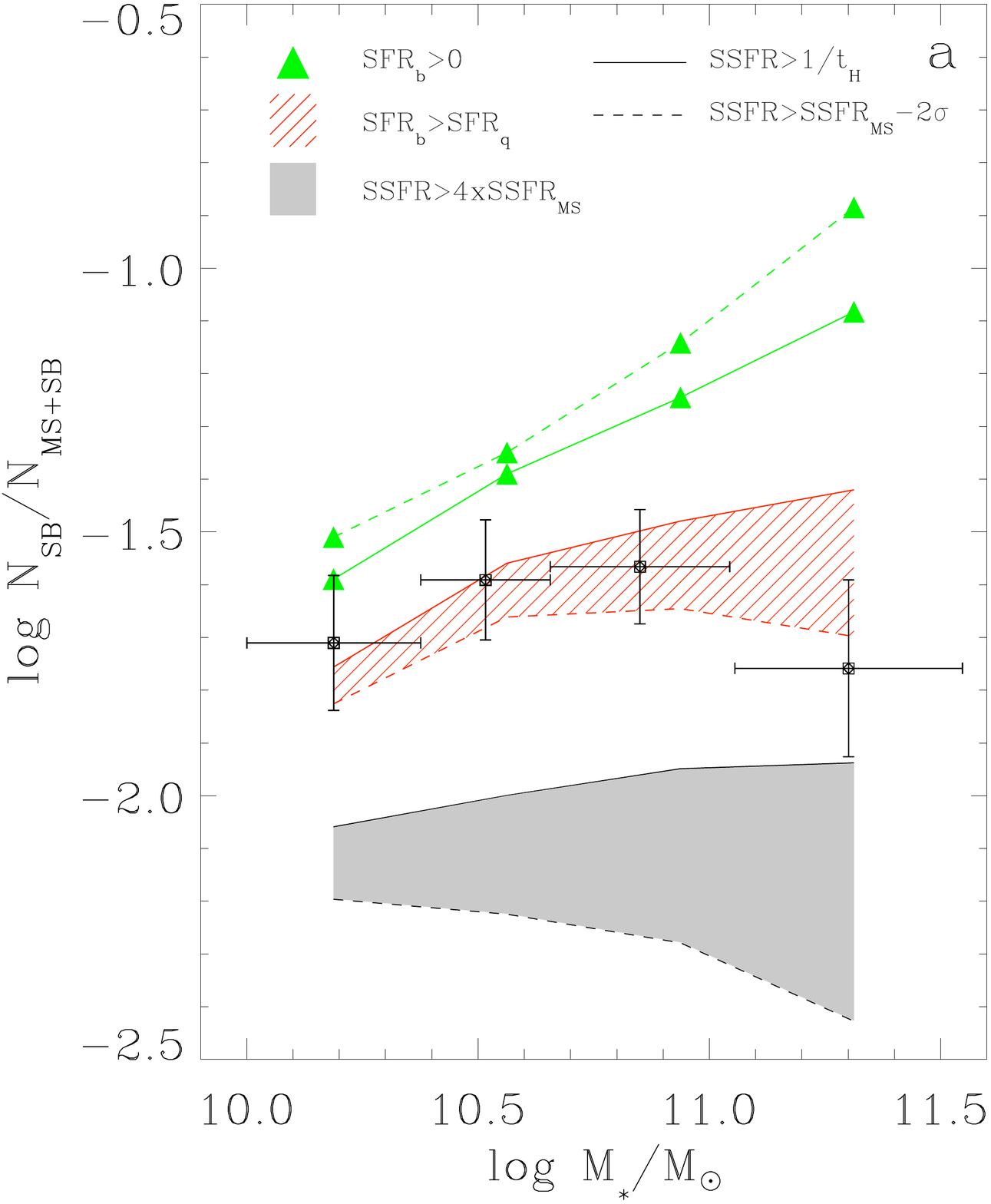}
\includegraphics[width=8cm]{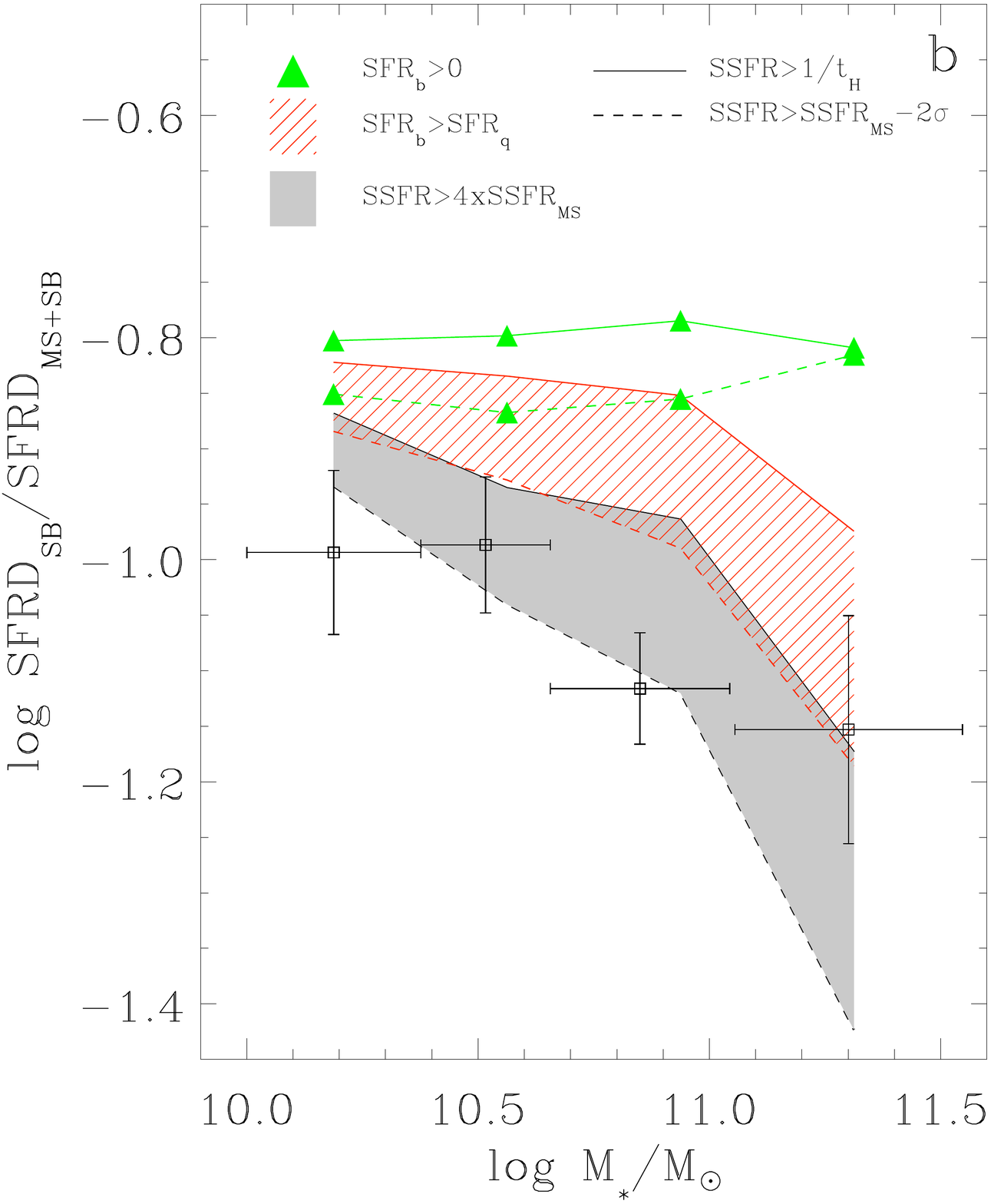}
\caption{Left: fraction of starburst (shaded region),  burst-dominated (hatched region), and  interacting (triangles)  galaxies  as function of the stellar mass at  $1.5<z<2.5$. Right: contribution to the cosmic SFRD  of the same  classes of starforming galaxies.  The lines indicate the model predictions adopting the  SSFR$ > t_H(z)^{-1}$ (solid lines)  and the ${\rm SSFR>SSFR_{MS}}$-2$\sigma$  (dashed lines) criterion to define the global sample of starforming galaxies. The data points indicate the results obtained by \cite{Rodighiero11}  for starburst galaxies  (SSFR$>$4$\times$ ${\rm SSFR_{MS}}$).}\label{sfrd}
\end{center} 
\end{figure*}

We now quantify the results shown in the previous section, and derive the relative contribution of starbursts and quiescently starforming galaxies to the cosmic SFRD and comoving number density in the redshift interval $1.5<z<2.5$. To perform a quantitative comparison we first
need to define the sample of model starforming galaxies. Since the observational results we compare with are derived from a combination of samples adopting different criteria to select starforming galaxies (see Sect. \ref{sfr_m}), we have performed two different selections of model galaxies in order to reproduce, at least partially, the observational selection effects.  The first corresponds to model galaxies with SSFR$ >$ ${\rm SSFR_{MS}}$-2$\sigma$ (see dashed line in fig. \ref{sfr_mst}b), where ${\rm SSFR_{MS}}$ is the value obtained for model galaxies and  $\sigma=0.3$ dex  is the scatter of the observed SFR-M$_*$ relation \citep{Rodighiero11,Daddi04}. The second corresponds to model galaxies with SSFR$ > t_H(z)^{-1}$ (see the thin solid line in fig. \ref{sfr_mst}b which indicates the  corresponding SSFR value at $z=2$), which select galaxies whose
current SFR  is stronger than its average past SFR \citep[see e.g.][]{Fontana09}.
Within such samples, we derive the distributions of SSFR normalized to the value of main sequence galaxies (${\rm SSFR_{MS}}$), in four equally spaced stellar mass bins from M$_*=10^{10} {\rm M_{\odot}}$ to M$_*=10^{11.5} {\rm M_{\odot}}$ (see fig. \ref{ssfr_dist}). According to \cite{Rodighiero11}, the galaxies observationally classified as starburst are those with SSFR$_{SB} > 4\times$ ${\rm SSFR_{MS}}$ (vertical dotted lines in fig. \ref{ssfr_dist}).
We then proceed to assess how galaxies observationally classified as starburst compare with model galaxies. To this aim, we derive separately the distributions of model galaxies dominated by the quiescent mode of star formation (blue histograms) and of galaxies dominated by the burst mode of star formation (red histograms). We find that the threshold SSFR$_{SB} > 4\times$ ${\rm SSFR_{MS}}$ 
used by \cite{Rodighiero11} is indeed effective in filtering out quiescent galaxies, independently of the criterion adopted to define
our sample of starforming galaxies; in fact, in the model SSFR values above this threshold are obtained only through the burst component of the star formation. However this criterion fails to select a considerable part of the burst-dominated starforming galaxies where the galaxy interactions do not boost the SFRs above the quiescent values (minor mergers and fly-by events).
The above results enlighten the physical difference between galaxies observationally classified as starbursts and galaxies with an interaction-driven star formation component. The latter also include galaxies not dominated by the burst mode of star formation, as it is indicated by the distributions of galaxies with a burst component of star formation (${\rm SFR_{b}}>0$, green histograms in fig. \ref{ssfr_dist}). In the following we will derive the relative contribution to the comoving number density and to the cosmic SFRD in the redshift range $1.5<z<2.5$ for all these classes of starforming galaxies: (i) main sequence galaxies (${\rm SFR_{q}}>{\rm SFR_{b}}$); (ii) starburst galaxies (${\rm SSFR_{SB}}$ $> 4\times$ ${\rm SSFR_{MS}}$), (iii) burst-dominated starforming galaxies (${\rm SFR_{b}}>{\rm SFR_{q}}$); and (iv) galaxies with an interaction-driven star formation component (${\rm SFR_{b}}>0$). \\
The predicted relative contribution of the above classes of galaxies to the comoving number density of starforming galaxies is shown in figure \ref{sfrd}a. The fraction of starbursts (shaded region) and burst-dominated galaxies (hatched region) with respect to
the total number of starforming galaxies is compared with the observational results by \cite{Rodighiero11}.
While the predicted fraction of starburst galaxies remains around ${\rm N_{SB}/N_{MS+SB}}$ $\sim 1\%$, the exact values predicted by the model depend on the criteria adopted to define the global sample of starforming galaxies (i.e., ${\rm N_{MS+SB}}$), defining the upper and the lower
envelope of both the shaded and the hatched regions. The smaller fraction of starburst galaxies obtained with the selection SSFR$ >$ ${\rm SSFR_{MS}}$-2$\sigma$ compared to the selection SSFR$ > t_H(z)^{-1}$ is due to the larger number of ``on-sequence'' galaxies selected in the first case (especially for large masses, see solid and dashed histograms in fig. \ref{ssfr_dist}), while the number of galaxies N$_{SB}$ remains almost unchanged. \\
The figure shows that, even in a hierarchical model connecting the physics of galaxies to their merging histories,
the fraction of galaxies classified as starburst remains within values $\sim 1\%$, even lower than the observational values ($\sim $2\%); values
around 2-3$\%$ are obtained when burst-dominated galaxies are considered.
This is due to a twofold reason: i) in hierarchical models, the statistics of merging events is dominated by minor episodes which do not produce
SSFR above the threshold adopted in the observations; ii) the duration of bursts $t\approx 10^7-10^8$ yrs is much smaller than the typical time scale
between two encounters, so that the probability to find a galaxy in a starburst phase is correspondingly small.
As for former point i), we show in figure \ref{sfrd}a the fraction of the galaxy population with a star formation component induced by galaxy interactions, ${\rm SFR_{b}}>0$ (triangles). When all galaxy interactions are considered, including minor merging and fly-by, the fraction of galaxies with an interaction-driven star formation component increases with M$_*$ to reach values $\approx 10^{-1}$. The trend with M$_*$ is due to the larger cross section for interactions of massive galaxies (see eq. \ref{int}) and to the fact that massive galaxies are found in dense environments with enhanced galaxy interaction rate. 
 The merging histories of massive galaxies are dominated by minor merging events which are less effective in destabilizing the gas in the galaxy disks (see eq. \ref{f}),  therefore a lower number of massive galaxies satisfy the ${\rm SFR_{b}}>{\rm SFR_{q}}$ criterion, and this explains the flattening of the hatched and shaded regions with increasing stellar mass in figure \ref{sfrd}a.

As for the starburst duty cycle, the above value $\sim 10^{-1}$ of the fraction of galaxies with ${\rm SFR_{b}}>0$ implies a value $\tau_b/\tau_r\sim 0.1$
for the ratio of the burst duration $\tau_b$ over the interaction time scale $\tau_r$ (see sect. 2.2). This is indeed what is expected from both theoretical
arguments and observational hints. In fact, the interaction time scale $\tau_r=(n\,\Sigma\,V)^{-1}$ can be estimated assuming a simple geometrical cross section $\Sigma\approx \pi r^2$. For typical values of the galaxy gravitational radius $r \sim 0.1$ Mpc and of the galaxy relative velocities $V\sim 100$ km/s, and assuming a density of a few galaxies per Mpc$^3$, we obtain an estimate $\tau_r\sim 10^9$ yr. When compared with the galaxy crossing time (determining the duration of bursts, see sect. 2.2) $\tau_b\sim 10^8$ yr, a value $\tau_b/\tau_r\approx 0.1$ is obtained for the duty cycle, as expected. This value is also consistent with the duty cycle of AGN measured at $z\approx 1-2$ with luminosity $L\gtrsim 10^{43}$ erg/s
\citep{Martini09, Eastman07}, as expected in our framework where both starbursts and AGNs are triggered by galaxy interactions.\\

\begin{figure}[h!]
\begin{center}
\includegraphics[width=7 cm]{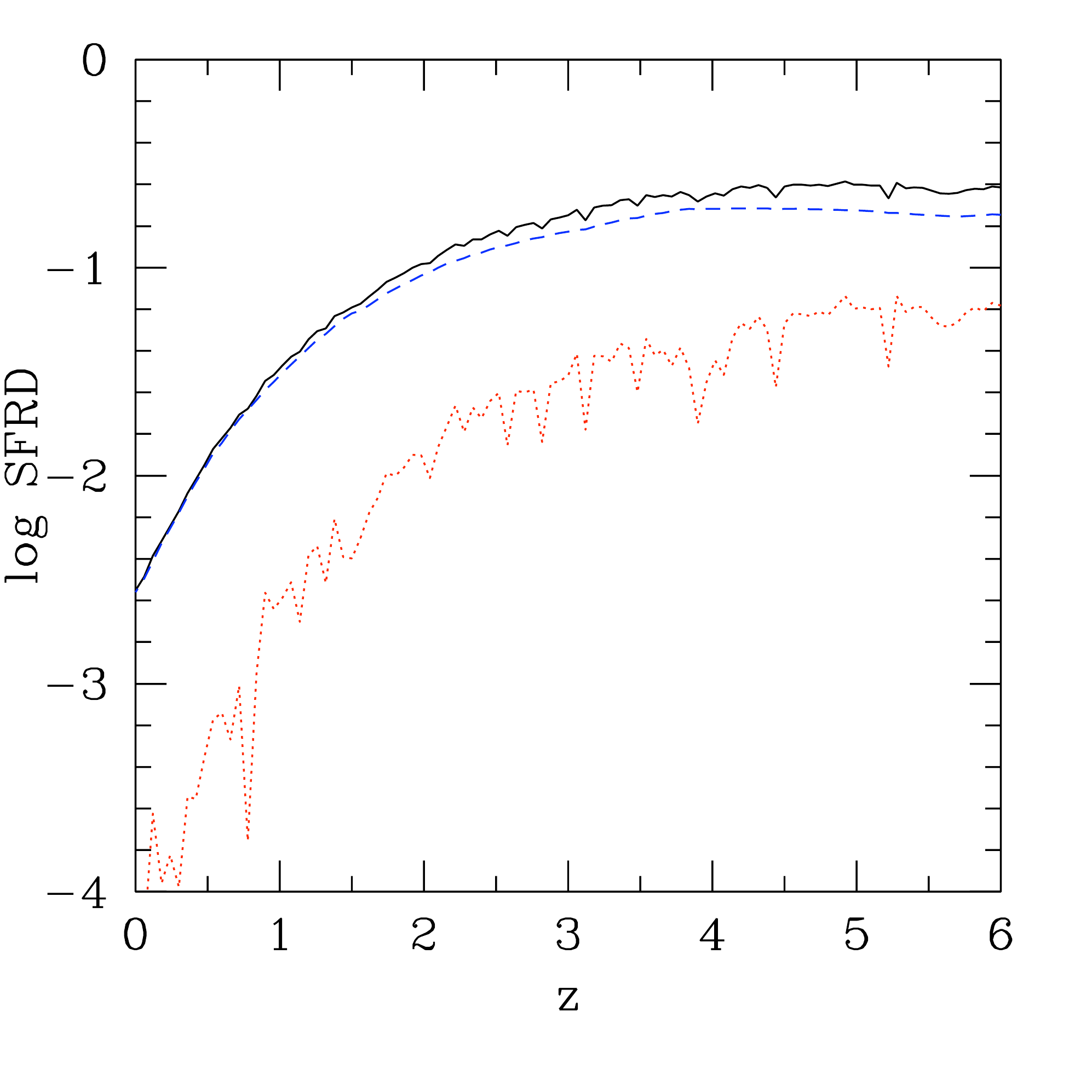}
\caption{Contribution to the cosmic SFRD of galaxies with SFR$ \geq $1.4$ \times 10^{-6}$ M$_{\odot}$/yr  (corresponding to UV luminosity L$_{UV} \gtrsim$ $ 10^{15} $ W Hz$^-1$ adopting the calibration given in \citealt{Kennicutt98}) as a function of redshift (solid line). Dashed line indicates the contribution of galaxies dominated by the quiescent mode of star formation, while dotted line indicates the contribution of galaxies dominated by the burst mode of star formation.}\label{sfrd_z}
\end{center} 
\end{figure}

\begin{figure}[h!]
\begin{center}
\includegraphics[width=7 cm]{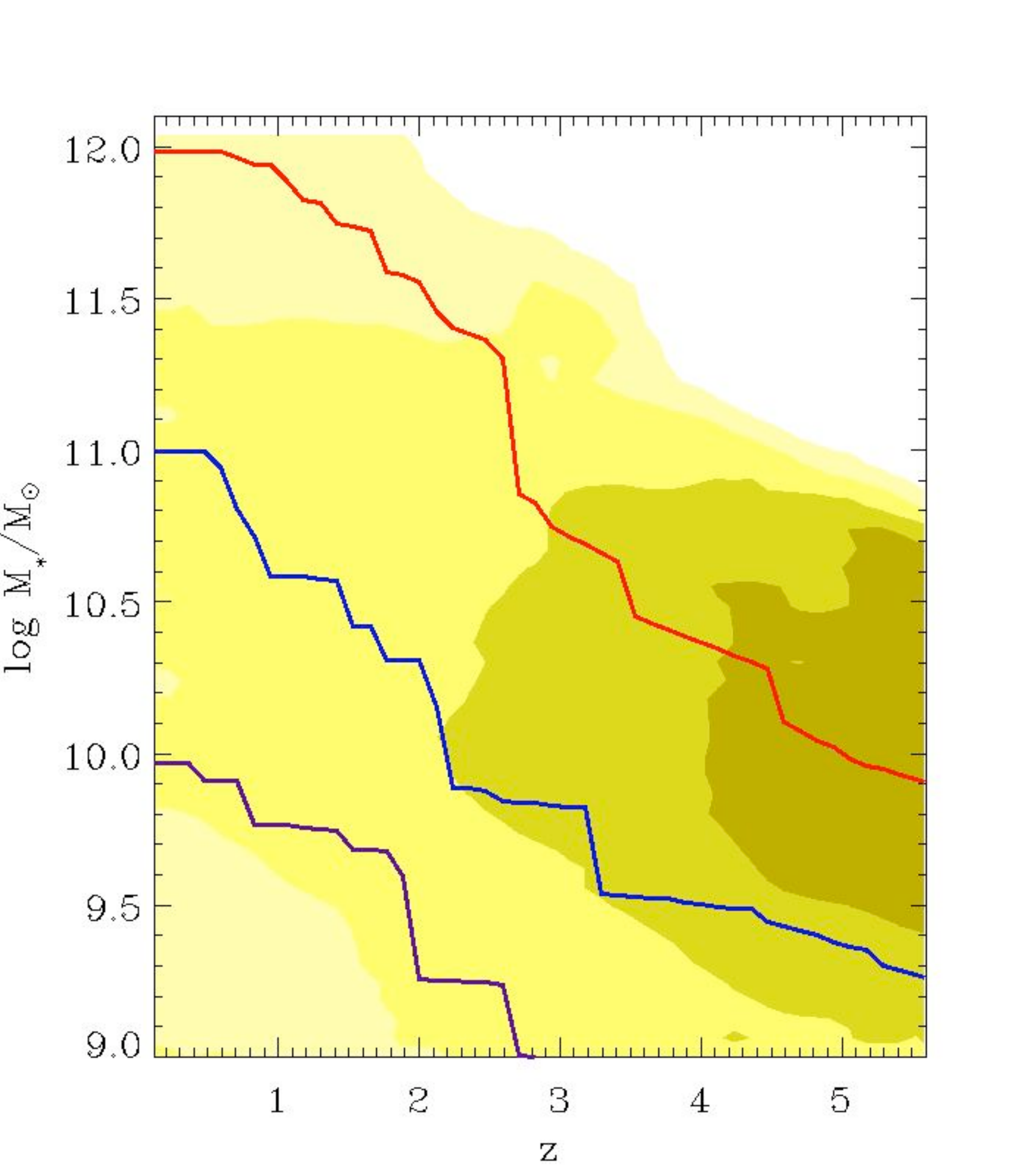}
\caption{The predicted average values of the ratio ${\rm M_{burst}(z)/M_*(z)}$ as a function of redshift and galaxy stellar mass. The four filled contours correspond to equally spaced values of the ratio  ${\rm M_{burst}(z)/M_*(z)}$: from 0.05  for the lightest filled region to  0.2 for the darkest. The lines illustrate the growth histories of a typical galaxy with final mass: M$_*(z=0)=10^{12} {\rm M_{\odot}}$ (red), M$_*(z=0)=10^{11} {\rm M_{\odot}}$ (blue), and M$_*(z=0)=10^{10} {\rm M_{\odot}}$ (purple).}\label{Mburst}
\end{center} 
\end{figure}

We now proceed to estimate
the contribution of the different classes of starforming galaxies to the cosmic SFRD at $1.5\leq z\leq 2.5$, shown in figure \ref{sfrd}b.
The interaction-driven scenario predicts that the contribution of the entire interacting galaxy population is $ \sim $15\% irrespective of stellar mass. At lower stellar masses this quantity is mainly contributed by burst-dominated galaxies, while at higher stellar masses the contribution of the quiescent mode of star formation becomes important as it is indicated by the decrease of the ratio ${\rm SFRD_{SB}/SFRD_{MS+SB}}$ with increasing M$_* $ for the ${\rm SFR_{b}}>{\rm SFR_{q}}$ and ${\rm SSFR_{SB}}$ $> 4\times$ ${\rm SSFR_{MS}}$ populations.
In fact, for the burst-dominated starforming galaxy population the contribution to the cosmic SFRD is $ \sim $15\% for M$_* \sim 10^{10} {\rm M_{\odot}}$ and (6-10)\% for M$_* \sim 10^{11.5} {\rm M_{\odot}}$, and for starburst galaxies it is slightly lower: $\sim$ 12\% for M$_* \sim 10^{10} {\rm M_{\odot}}$ and (4 -6)\% for M$_* \sim 10^{11.5} {\rm M_{\odot}}$. Therefore, this scenario is consistent with the observational finding that the cosmic SFRD in the redshift range 1.5$<z<$2.5 is mainly contributed by quiescently starforming galaxies  \citep{Rodighiero11}. However, it must to be noted that, given the lower fraction of starburst galaxies predicted by the model, the match between the predicted and observed starburst contributions is determined by the lower value of the SFR of model main sequence galaxies (see figure \ref{sfr_mst}). \\
Since the burst mode of star formation is intimately related to the galaxy interaction rate, which is expected to increase with redshift, we investigate how these contributions change as a function of the cosmic epoch. To address this issue, we derive separately the contribution to the cosmic SFRD of quiescent-dominated and burst-dominated systems as a function of redshift (dashed and dotted lines in figure \ref{sfrd_z} respectively). We find that quiescent systems dominate the global SFRD at all redshifts. However, the contribution of the burst-dominated population increases with redshift, rising from $\lesssim$ 5\% at low redshift ($ z\lesssim 0.1$) to $ \sim $ 20\% at $ z\gtrsim 5$. This is a typical feature of the hierarchical clustering scenario where the starburst events are triggered by galaxy interaction during their merging histories; a similar evolution was obtained by
\cite{Hopkins10} on the basis of cosmological hydrodinamical simulations.
The physical origin of this behaviour can be understood as follows: at high redshift both the interaction rate $\tau_r^{-1}$ (eq. \ref{int}) and the fraction of destabilized gas $f$ (eq. \ref{f}) are large; the former due to the large densities, the latter due to the large ratio $m'/m\approx 1$ characteristic of this early phase ($z\gtrsim 3$) when galaxy interactions mainly involve partners with comparable mass (major mergers). At lower $z$ the decline of the interaction rate and of the destabilized gas fraction suppresses the burst mode of star formation thus lowering its contribution to the total SFRD.\\
An immediate implication of the above is that the fraction of the stellar mass formed through the SFR associated with the destabilized cold gas during galaxy interactions increases with redshift. This is illustrated in figure \ref{Mburst}, where the contours show the average values of the ratio ${\rm M_{burst}(z)/M_*(z)}$ as a function of M$_*$ and $z$. The mass ${\rm M_{burst}(z)}$ is the stellar mass that a galaxy has formed through the burst mode of star formation up to that redshift, and M$_*(z)$ is the total galaxy stellar mass at that $ z $. From this figure one can infer that $ \sim $10\% of the final ($ z=0$) galaxy stellar mass has been formed during bursts for $10^{10} {\rm M_{\odot}}\leq$M$_*\leq10^{11.5} {\rm M_{\odot}}$. At higher redshift ($ z \gtrsim 4$) this fraction increases up to $ \sim $20\%.
To illustrate how the ratio ${\rm M_{burst}(z)/M_*(z)}$ varies during the formation of galaxies with different final masses, we also show the growth histories of typical galaxies with final mass: M$_*(z=0)=10^{12} {\rm M_{\odot}}$, M$_*(z=0)=10^{11} {\rm M_{\odot}}$, and M$_*(z=0)=10^{10} {\rm M_{\odot}}$ (red, blue, and purple lines in fig. \ref{Mburst}).
These paths illustrate that at high redshift ($ z\gtrsim 2$) and for massive objects large values of   ${\rm M_{burst}(z)/M_*(z)}$ are expected, since such massive galaxies formed in biased regions of the density field where high-redshift interactions are extremely effective in triggering starbursts. For less massive galaxies the high-redshift value of the ${\rm M_{burst}(z)/M_*(z)}$ ratio progressively lowers, since these galaxies formed in less dense environments.\\

\section{Discussion }\label{discussion}
In this section we discuss the origin of the mismatch between the normalization of the predicted and observed galaxy main sequences which was highlighted in section \ref{sfr_m}. 
Similar mismatch  between the observed SFRs and those expected by various kind of SAMs and hydrodinamical simulations  has been also reported by several authors at $z \lesssim 2$  \citep[see e.g.][]{Daddi07, Dave08, Fontanot09, Damen09, Santini09,Lin12}. 
This discrepancy either indicates an incompleteness in our knowledge of galaxy assembly, or systematic effects in stellar mass and/or SFR determinations, or both.\\
On the observational side, the uncertainties in stellar mass and SFR estimates may be responsible for one of the main puzzles that appear in present-day observational cosmology: the mismatch between the observed stellar mass density  and the integrated SFRD  \citep[e.g.][]{Hopkins06, Fardal07, Wilkins08, Santini12}.
In principle, these two observables represent independent approaches to studying the mass assembly history from different points of view. However, the integrated SFRD, after considering the gas recycled fraction into the interstellar medium, appears to be higher than the observed stellar mass density by a factor of about 2-3 at $ z \leq 2$   \citep{Santini12}. Moreover, if the merging contribution to the stellar mass build-up is accounted for  \citep{Drory08}  the agreement gets even worse. Intriguingly, the integrated SFRD exceeds the observed stellar mass density, implying that either the SFR are overestimated, or the stellar masses are underestimated. Both effects would reduce the offset between the observed and predicted SFR-M$_*$ relations at least by a factor of 2.\\
On the theoretical side, we have investigated whether the mismatch can be ascribed to the implementation of the star formation law usually adopted in SAMs. In this respect, the most critical point is the assumption that the star formation law in eq. (\ref{sfr_q}) - derived from the local observed Kennicutt relation \citep{Kennicutt98} - is valid at any redshift. To investigate the effect of relaxing this assumption, we have varied the high-redshift star formation timescale (the
free parameter $q$ in eq. \ref{sfr_q}) so as to obtain star formation efficiencies SFR$/m_c$ above the local value for $z\geq 2$. However, even increasing the efficiency up to a factor of 4, we do not find an appreciable up-shift of the model main sequence. In fact, increasing the high-$z$ star formation results into an earlier consumption of the the cold gas reservoir of model galaxies, an effect which balances the increased star formation efficiency leaving the normalization of the SFR-M$_*$ relation almost unchanged.\\
 It is also interesting to examine the impact of other  star formation laws on the shape and normalization of the main sequence. This was investigated by \cite{Lagos11b} by implementing the empirical relation of \cite{Blitz06} and the theoretical model of \cite{Krumholz09} in the SAMs of  \cite{Baugh05} and \cite{Bower06}. These star formation laws are of the form $\Sigma{\rm_{SFR}}=\nu_{SF}\Sigma_{mol}=\nu_{SF}f_{mol}\Sigma_{gas}$, where $\Sigma{\rm_{SFR}}$ and  $\Sigma_{gas}$ are the surface densities of the SFR and total cold gas mass, $ \nu_{SF} $  is the inverse of the star formation time scale for the molecular gas, and  $f_{mol}=\Sigma_{mol}/\Sigma_{gas}$  is the molecular-to-total gas mass surface density ratio which depends on the hydrostatic gas pressure \citep{Blitz06} or on the balance between the dissociation of molecules due to the interstellar far-UV radiation and their formation on the surface of dust grain \citep{Krumholz09}.
They found that the main sequence appears insensitive to the star formation law. 
 Thus, tuning the quiescent star formation law does not seem to provide a solution to the mismatch between the normalization of the predicted and observed  galaxy main sequence. It is however possible that other ways of modelling starburts, e. g. starburst driven by disk instabilities \citep[e. g.][]{Bower06,Lagos11b,Guo11,Fanidakis11a,Fanidakis11b,Hirschmann12}, could have an impact on the shape and normalization of the main sequence \cite[see][]{Lagos11b}.  While the latter constitutes an interesting complement to  the merger-driven starbursts, its consistence  with the observed fraction of starbursts is still to be investigated.

\begin{figure}[h!]
\begin{center}
\includegraphics[width=7cm]{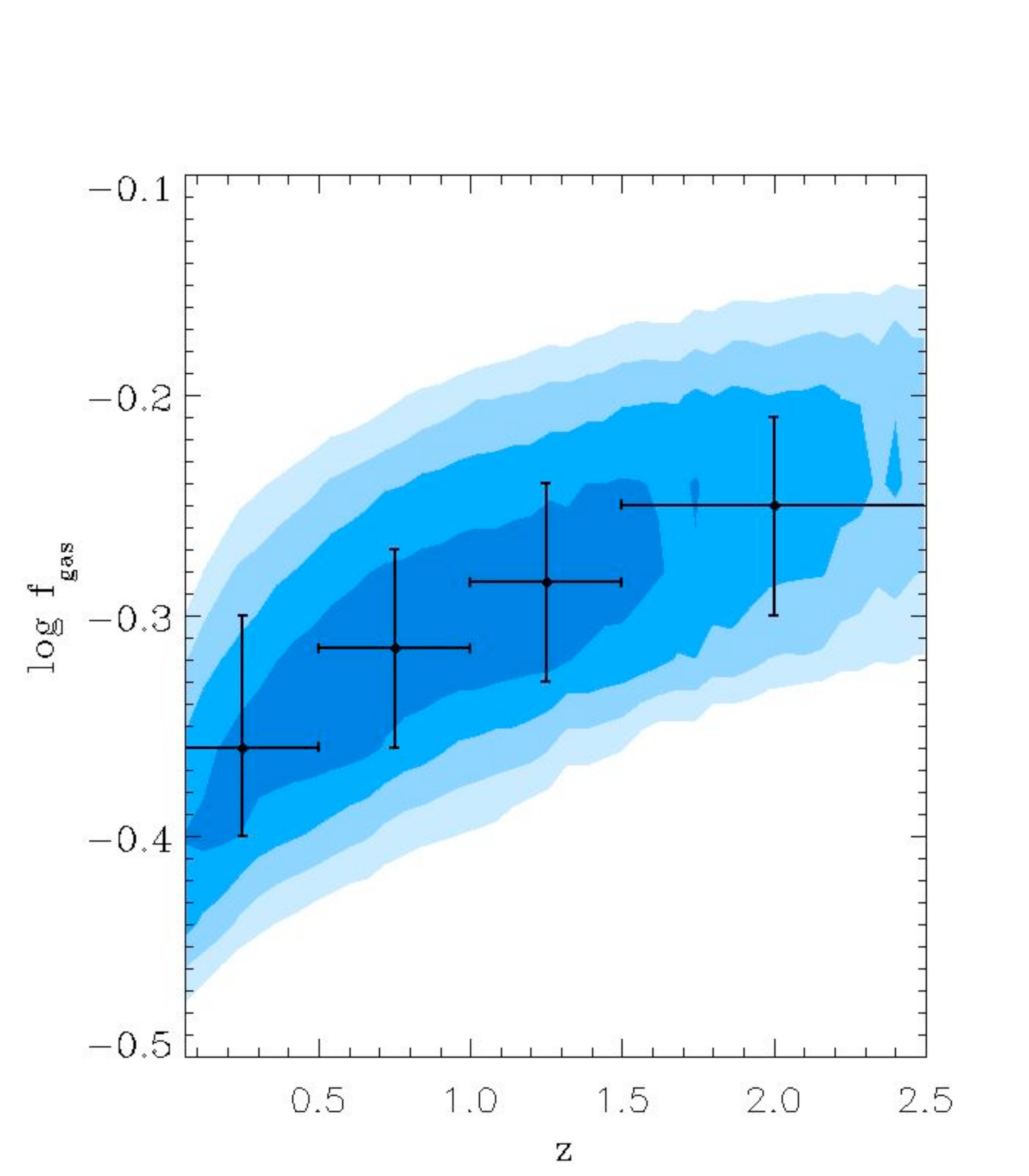}
\caption{Gas fraction distributions as a function of redshift of model galaxies with SSFR$>t_H(z)^{-1}$  and M$_*$ $> 10^{10} {\rm M_{\odot}}$. The  filled contours correspond to equally spaced values of the density (per Mpc$^3$) of galaxies in logarithmic scale: from 10$^{-3}$  for the lightest filled region to  10$^{-2.1}$ for the darkest. The plotted value of log $f_{gas}$ is the median value for sources in four  redshift bins: 
  log $f_{gas}$=$-0.36_{-0.04}^{+0.06}$ (0$<z<$0.5);   log $f_{gas}$=$-0.32_{-0.05}^{+0.05}$ (0.5$<z<$1);  log $f_{gas}$=$-0.29_{-0.05}^{+0.05}$ (1$<z<$1.5);   log$f_{gas}$=$-0.25_{-0.05}^{+0.04}$ (1.5$<z<$2.5). The vertical error bars show the interpercentile range  containing 68\% of the sources.}\label{fgas}
\end{center} 
\end{figure}

An alternative possibility is that the cold gas fractions predicted by the models at $z\approx 2$ underestimates the real values, due to a too early conversion of cold gas into stars at earlier cosmic times. However, limiting the cooling and star formation efficiency at high redshifts is an extremely challenging task in the hierarchical clustering scenarios. These predict the high redshifts progenitor galaxies to be characterized by low virial temperature and high densities, making the gas cooling extremely efficient at high redshifts $z\geq 3$; the short dynamical timescales and the frequent merging events rapidly convert such available cold gas into stars, leaving only a residual fraction of galactic baryons in the form of cooled gas available for star formation at $z\approx 2$. Although such an over-cooling problem \citep[already pointed out in the seminal paper by][]{White78} is alleviated by stellar feedback expelling part of the cold gas from the starforming galactic regions, the efficiency of the latter process is severely limited at high redshifts due to the compactness of dark matter haloes resulting in large escape velocities. In fact, such a long-standing problem of the hierarchical scenario of galaxy formation leads to the over-prediction of low-mass galaxies  at $ z\geq 1.5$, as shown by the comparison of our and other SAMs with the observed stellar mass function \citep{Fontana06, Fontanot09, Marchesini09, Guo11, Santini12}.
The excess of starforming low-mass galaxies at $ z\geq 1.5$ reflects into an excess of red low-mass galaxies at $ z\sim 0$, even when a large feedback is implemented in the models \citep{Guo11}.\\ 
Assessing whether the evolution of the cold gas in galaxies is the key to explain (at least partially) the SAMs under-prediction of SSFRs at $z\leq 2$ 
would require a comparison between the predicted and the observed galaxy cold gas fraction as a function of redshift.  The evolution of the cold gas content of galaxies represents also one way of testing time-changing star formation laws, since the latter  determine the rate at which gas is converted into stars. An increase in the star formation time scale leads to larger cold gas masses in disks and vice versa \citep[see e.g. ][]{Lagos11c,Lagos11b,Fu12}. In models where starbursts are triggered  by disk instabilities, a longer star formation time scale in disks can lead to lower final gas content in some cases, being the disk more prone to instabilities if the total mass is larger \citep{Lagos11b}. 
The predictions for such quantity in our model are shown in figure \ref{fgas} in terms of the gas fraction parameter $f_{gas}=M_{gas}/(M_{gas}+M_{*})$, for model galaxies with SSFR$>t_H(z)^{-1}$ and M$_*$ $> 10^{10} {\rm M_{\odot}}$ , as predictions to be tested with future observations.\\


It is also to be noted that the offset between the observed and predicted galaxy main sequence can be due not only to an underestimation of the predicted SFRs but also to an overestimation  of the predicted stellar masses. A physical process that can reduce the stellar mass of galaxies is the disruption of stellar material from merging satellites due to  tidal stripping \citep[e.g.][]{Henriques10}. Evidence for the importance of this process in galaxy formation comes from  the existence of a diffuse population of intra-cluster stars  \citep{Zwicky51, Durrell02}. Indeed, gas-dynamical simulations \citep[e.g.][]{Moore96} generally agree that  these stars have been continually stripped from member galaxies throughout the lifetime of a cluster, or have been ejected into intergalactic space by merging galaxy groups , rather than having formed in the intra-cluster medium.

\section{Conclusions}\label{conclusion}
We have investigated the relative importance of the quiescent mode and burst mode of star formation in determining the SFR-M$_*$ relation and the evolution of the cosmic SFRD, predicted under the assumption that starburst events are triggered by galaxy encounters during their merging histories. The latter are described through Monte Carlo realizations, and are connected to gas processes and star formation using a SAM of galaxy formation in a cosmological framework. The main results of this paper are:
\begin{itemize}

\item Hierarchical clustering models reproduce the slope and the scatter of the SFR-M$_*$ relation, however, they under-predict  the normalization by a factor of 4-5 a $z \simeq 2$ \citep{Daddi07, Dave08, Fontanot09, Damen09, Santini09, Lin12}.  A possible theoretical explanation of this mismatch is that the amount of cold gas in  galaxy disks predicted by these models at $ z\simeq 2$ underestimate the real values.  We derive the prediction for the evolution of the gas fraction distribution of starforming galaxies, as prediction to be tested with future observations. 

\item  The tight correlation between SFR and  M$_*$ of the galaxies on the main sequence is determined by the quiescent component of the star formation.  Galaxies dominated by the star formation component induced by galaxy interactions populate the regions well above (major mergers) and below (minor mergers and fly-by events) the sequence. 

\item The predicted SSFR distributions of starforming galaxies indicate that galaxies observationally classified as starburst on the basis of their distance from the main sequence (${\rm SSFR_{SB} > 4\times SSFR_{MS} }$),  are indeed dominated by the burst component of the star formation. However, this criterion fails to select part of the burst-dominated starforming galaxies where galaxy interactions do not strongly boost the star formation rate.
 
 \item Hierarchical clustering scenarios, connecting the properties of galaxies to their merging histories, naturally yield fraction of starburst galaxies at $ z\sim 2$  of $ \sim $1\%, in agreement with the observational estimate \citep[2\%][]{Rodighiero11}. This low fraction is due to the fact that in this scenario, the statistic of merging events is dominated by minor episodes which do not produce SSFR above the observational threshold, and to the small probability of finding a galaxy in a starburst phase due to the small value of the duration of the burst ($t\approx 10^7-10^8$) yrs compared with the typical time scale between galaxy encounters.   

\item  Quiescently starforming galaxies dominate the global SFRD at all redshifts. The contribution of the burst-dominated systems increases with redshift, rising from $\lesssim$ 5 \% at $ z\lesssim 0.1$ to $ \sim $ 20\% at $ z \geq 5$. This behaviour is determined by the increase of the  galaxy interaction rates and of the effectiveness of galaxy interaction in destabilizing the cold gas at high redshift. The fraction of the final ($ z =0$) galaxy stellar mass  which is formed through the burst component of star formation is  $ \sim $10\% for 
$10^{10} {\rm M_{\odot}} \leq$M$_*\leq10^{11.5} {\rm M_{\odot}}$.




\end{itemize}
These findings do not imply that  galaxy mergers/interactions play a lesser role in the formation of stars in galaxies. Indeed, for each galaxy the star formation (both the quiescent and burst component) is driven by the cooling rate of the hot gas and by the rate of refueling of the cold gas, which in turn is intimately related to the galaxy merging histories.\\ 
We found that the interaction-driven model for  starburst galaxies provides strict predictions  for the evolution and the mass dependence of the starburst contribution to the cosmic SFRD, which represent effective tools to test this scenario with future observations of large and complete samples of starforming galaxies at $ z > 2$.
To reach this goal, it is also necessary a reliable observational diagnostic which is able to distinguish between galaxies dominated by the quiescent and burst mode of star formation. The measure of the star formation efficiency and of the IR star formation compactness  could represent eligible  candidates for this purpose \citep{Daddi10, Genzel10, Elbaz11}. The first is based on measurements of the mass and density of molecular gas at high redshift, and  on the poor knowledge  of the CO luminosity to H$_2 $ conversion factor.  The technique of separating  starbursts and quiescently starforming galaxies based on their IR star formation compactness seems to be promising, since ALMA will provide powerful tools to measure the spatial distribution of star formation in distant galaxies at high angular resolution, making it possible  to estimate the compactness  of the star formation regions.

\section*{Acknowledgments} The authors thank Giulia Rodighiero for kindly providing the data plotted in figure 2, and the referee for helpful comments. This work was supported by ASI/INAF contracts I/024/05/0 and I/009/10/0 and PRIN INAF 2011.  \ref{sfr_mst}.

\bibliographystyle{aa}
\bibliography{biblio.bib}

\end{document}